\documentstyle[aps,epsf]{revtex}  

\def\ra{\rightarrow}
\def\be{\begin{equation}}
\def\ee{\end{equation}}
\def\bea{\begin{eqnarray}}
\def\eea{\end{eqnarray}}

\def\pieta{$\pi^0$-$\eta,\eta'\;$}
\def\Bpipi{$B \ra \pi\pi\;$}

\def\lapp{\hbox{$ {     \lower.40ex\hbox{$<$}
                   \atop \raise.20ex\hbox{$\sim$}
                   }     $}  }
\def\rapp{\hbox{$ {     \lower.40ex\hbox{$>$}
                   \atop \raise.20ex\hbox{$\sim$}
                   }     $}  }
\def\hhha{\rule[-3.mm]{0.mm}{7.mm}}

\begin{document}
\baselineskip 14pt 
\title{
\begin{flushright} \normalsize
UK/TP 99-08 \\  
hep-ph/9906269 \\
\end{flushright}
Isospin Violation in ${\bf B\ra \pi\pi}$ Decays
\footnote{Talk at American Physical Society, Division of Particles and Fields Conference, Los Angeles, CA, Jan. 5-9, 1999.}}

\author{S. Gardner}

\address{Department of Physics and Astronomy, University of Kentucky, \\
Lexington, KY 40506-0055 \\E-mail: gardner@pa.uky.edu} 


\maketitle
\begin{abstract}
An isospin
analysis of $B\rightarrow \pi\pi$ decays 
yields $\sin 2\alpha$, where $\alpha$ is the usual CKM angle
$\alpha\equiv {\rm arg} [-V_{td} V_{tb}^\ast/(V_{ud}V_{ub}^\ast)]$
without hadronic uncertainty if isospin is a perfect symmetry.
Yet isospin symmetry is broken 
not only 
by electroweak effects but also 
by the $u$ and $d$ quark mass difference --- 
the latter drives
$\pi^0-\eta,\eta'$ mixing and 
converts the isospin-perfect triangle relation between the $B\ra \pi\pi$
amplitudes to a quadrilateral. 
The error incurred in $\sin 2\alpha$ through the neglect of the
resulting isospin-violating effects can be significant, 
particularly if $\sin 2\alpha$ is small.
\end{abstract}

%
%

\section{Introduction}

In the standard model, 
CP violation is characterized by a single phase in the 
Cabibbo-Kobayashi-Maskawa (CKM) matrix, rendering its elements complex. 
The CKM matrix is also unitary, so that 
determining whether or not this is empirically so is a central test of the
standard model's veracity~\cite{flrev96}. Ascertaining whether
the angles of the unitarity triangle, 
$\alpha$, $\beta$, and $\gamma$, empirically sum to $\pi$
and whether its angles are compatible with the measured lengths of
its sides lie at the heart of these tests of the standard model. 

We study the impact of isospin violation on the 
extraction of $\sin 2\alpha$ from an isospin analysis 
in \Bpipi decays~\cite{GL90}. Isospin is broken not only by 
electroweak effects but also by 
the $u$ and $d$ quark mass difference. The latter drives
$\pi^0-\eta,\eta'$ mixing~\cite{leut96}, which, in turn, generates an
amplitude in \Bpipi not included in the isospin analysis. Thus, although 
the effect of electroweak penguins is estimated
to be small~\cite{deshe95,gronau95,fleischer96}, 
when all the effects of isospin violation are included, 
the error in the extracted value of $\sin 2\alpha$ can 
be significant~\cite{pieta98}. 

  To review the isospin analysis
in $B\ra \pi\pi$ decays, due to Gronau and London~\cite{GL90}, 
let us consider 
the time-dependent asymmetry $A(t)$~\cite{pdg98}:
\be
A(t) = {( 1 - |r_{f_{CP}}|^2) \over ( 1 + | r_{f_{CP}}|^2)}
\cos(\Delta m\, t) 
- {2 ({\rm Im}\, r_{f_{CP}}) 
\over 
( 1 + | r_{f_{CP}}|^2)}
\sin (\Delta m\, t) \;, 
\ee
where $r_{f_{CP}} = ({V_{tb}^\ast V_{td} / V_{tb}V_{td}^\ast})
({{\overline A}_{f_{CP}}/ A_{f_{CP}}}) \equiv 
e^{-2i\phi_m} {{\overline A}_{f_{CP}} / A_{f_{CP}}} $,
$A_{f_{CP}}\equiv A(B_d^0\ra f_{CP})$, and 
$\Delta m\equiv B_H - B_L$~\cite{GKN97}.

Denoting the
amplitudes  $B^+ \ra \pi^+ \pi^0$, 
$B^0 \ra \pi^0 \pi^0$, and $B^0 \ra \pi^+ \pi^-$
by $A^{+0}$, $A^{00}$, and $A^{+-}$, respectively, 
and introducing $A_I$ to denote an
amplitude of final-state isospin $I$, we have~\cite{GL90}
\be 
{1\over 2} A^{+-} = A_2 - A_0 \;\;;\;\;  A^{00} = 2A_2 + A_0 \;\;; \;\;
{1\over \sqrt{2}} A^{+0} =3 A_2 \;,
\label{tridef}
\ee
where analogous relations exist for 
$A^{-0}$, ${\overline A}^{00}$, and ${\overline A}^{+-}$
in terms of ${\overline A}_2$ and ${\overline A}_0$. If isospin
were perfect, then 
the Bose symmetry of the $J=0$ $\pi\pi$ state would permit amplitudes merely
of $I=0,2$, so that 
the amplitudes $B^\pm \ra \pi^\pm \pi^0$ would be purely $I=2$.
In this limit 
the penguin contributions are strictly 
of $\Delta I=1/2$ character, so that they cannot contribute to the
$I=2$ amplitude: no CP violation is possible in the
$\pi^\pm\pi^0$ final states. The penguin contribution
in $B^0 \ra \pi^+ \pi^-$, or in 
${\bar B^0} \ra \pi^+ \pi^-$, 
can then be isolated and
removed by determining the relative magnitude and phase of the 
$I=0$ to $I=2$ amplitudes. We have 
\be
r_{\pi^+\pi^-}= e^{-2i\phi_m} {({\overline A}_2 - {\overline A}_0)
\over
({A}_2 - {A}_0)} = e^{2i\alpha} {(1 - {\overline z})\over (1 - z)}\;,
\label{rdef}
\ee
where $z ({\overline z}) \equiv A_0/A_2 ({\overline A}_0/{\overline A}_2)$
and ${\overline A}_2/A_2 \equiv \exp(-2i \phi_t)$ with
$\phi_t \equiv {\rm arg} ( V_{ud} V_{ub}^\ast)$ and 
$\phi_m + \phi_t = \beta + \gamma = \pi - \alpha$ in the 
standard model~\cite{pdg98}.  
Given 
$|A^{+-}|$, $|A^{00}|$, 
$|A^{+0}|$, and their charge conjugates,
the measurement of ${\rm Im}\, r_{\pi^+\pi^-}$ determines $\sin 2\alpha$,
modulo discrete ambiguities in ${\rm arg} ((1-{\overline z})/(1-z))$,
which correspond geometrically to the orientation  
of the ``triangle'' of amplitudes associated with Eq.~(\ref{tridef}),
namely 
\be 
 A^{+-} + 2A^{00} = \sqrt{2} A^{+0} \;,
\label{triangle}
\ee
with respect to $|A^{+0}|=|A^{-0}|$
and that
of its charge conjugate. 
The triangles' relative orientation can be resolved 
via a measurement of 
${\rm Im}\, r_{\pi^0\pi^0}$ as well~\cite{GL90}, and thus $\sin 2 \alpha$
is determined uniquely.  

\section{Isospin Violation and \pieta Mixing}

We examine the manner in which
isospin-violating effects impact the extraction of 
$\sin 2\alpha$, for isospin 
is merely an approximate symmetry. 
The charge difference
between the $u$ and $d$ quarks engenders a $\Delta I=3/2$
electroweak penguin contribution, which is outside the scope of the
delineated isospin analysis~\cite{GL90}, although methods have been 
suggested to include them~\cite{gronau98,pirjol99}.
This is the only manner in which the $u$-$d$ charge
difference enters our analysis, so that we term this source of isospin
breaking an ``electroweak effect.'' 
The $u$-$d$ quark mass difference can also engender a 
$\Delta I=3/2$
strong penguin contribution through isospin-breaking
in the hadronic matrix elements. 
Moreover, strong-interaction isospin violation
drives 
$\pi^0-\eta,\eta'$ mixing~\cite{leut96}, admitting an $I=1$ 
amplitude. Although electroweak penguin contributions
are estimated 
to be small~\cite{deshe95,gronau95,fleischer96}, 
other isospin-violating 
effects, such as \pieta mixing, 
can be important~\cite{pieta98,epsprime}. 

To include the effects of \pieta mixing, we write the
pion mass eigenstate $|\pi^0\rangle$ 
in terms of the SU(3)$_f$-perfect states $|\phi_3\rangle$, $|\phi_8\rangle$,
and $|\phi_0\rangle$, where, in the quark model,
$|\phi_3\rangle=|u{\overline u} - d{\overline d}\rangle/\sqrt{2}$, 
$|\phi_8\rangle=|u{\overline u} +
 d{\overline d} - 2s{\overline s}\rangle/\sqrt{6}$, 
and 
$|\phi_0\rangle=
|u{\overline u} + d{\overline d} + s{\overline s}\rangle/\sqrt{3}$. 
Explicit relations between the physical and SU(3)$_f$-perfect states can be 
realized by
expanding QCD to leading order in $1/N_c$, 
momenta, and quark masses to 
yield a low-energy, effective Lagrangian in which 
the pseudoscalar meson octet and singlet states are 
treated on the same footing~\cite{leut96,DGH92}. 
Diagonalizing the
quadratic terms in $\phi_3$, $\phi_8$, and $\phi_0$ of the
resulting effective Lagrangian
determines 
the mass eigenstates $\pi^0$, $\eta$,
and $\eta'$ and yields, 
to leading order in isospin violation~\cite{leut96},
\be
|\pi^0\rangle  = |\phi_3\rangle + \varepsilon 
(\cos \theta |\phi_8\rangle - \sin \theta |\phi_0\rangle)
+ \varepsilon' (\sin \theta |\phi_8\rangle + \cos \theta
|\phi_0\rangle) \;,
\label{physpi}
\ee
where $\cos \theta |\phi_8\rangle - \sin \theta |\phi_0\rangle
= |\eta\rangle + O(\varepsilon)$, 
and $\sin \theta |\phi_8\rangle + 
\cos \theta |\phi_0\rangle = |\eta'\rangle + O(\varepsilon')$. Moreover,
$\varepsilon = \epsilon_0 \chi \cos \theta$ and 
$\varepsilon' = -2\epsilon_0 \tilde\chi  \sin \theta$,
with $\chi= 1 + (4m_K^2 - 3m_{\eta}^2 - m_{\pi}^2)/(m_\eta^2 - m_\pi^2)
\approx 1.23$, $\tilde\chi = 1/\chi$, 
$\epsilon_0 \equiv \sqrt{3}(m_d - m_u) /(4(m_s - \hat{m}))$,
 and $\hat{m}\equiv(m_u + m_d)/2$~\cite{leut96}. Thus the magnitude
of isospin breaking is controlled by the SU(3)-breaking parameter 
$m_s - \hat{m}$. The $\eta$-$\eta'$ mixing angle $\theta$ 
is found to be
$\sin 2\theta= - (4\sqrt{2}/3)(m_K^2 - m_\pi^2)/(m_{\eta'}^2 - m_\eta^2)$
so that $\theta \approx -22^{\circ}$~\cite{leut96}. 
The resulting $\varepsilon
=1.14\epsilon_0$ is comparable
to the one-loop-order chiral perturbation theory 
result of $\varepsilon=1.23\epsilon_0$
in $\eta\ra \pi^+ \pi^- \pi^0$~\cite{gasser85a,leut96}. 
Empirical constraints also exist on the {\it sign} of \pieta
mixing. That is, the ratio of the reduced matrix elements in $K_{l3}$ decays,
namely, $K^+ \ra \pi^0 e^+ \nu_e$ and $K_L^0 \ra \pi^- e^+ \nu_e$, is
given by~\cite{na97}
\be
\left(\frac{f_+^{K^+\pi^0}}{f_+^{K_L^0\pi^-}}
\right)^{\rm expt} = 1.029 \pm 0.010 \;.
\ee
Using the Lagrangian of Ref.~\cite{leut96} and the quark masses 
$m_q(\mu=1\; {\rm GeV})$ of Ref.~\cite{ali98} yields
\be
\left(\frac{f_+^{K^+\pi^0}}{f_+^{K_L^0\pi^-}}
\right) = 1 + \sqrt{3} \varepsilon_8 \approx 1.018 \pm 0.010 \;, 
\ee
where $\varepsilon_8$ is the $\phi_3-\phi_8$ mixing angle, 
$\varepsilon_8\equiv \varepsilon \cos\theta + \varepsilon' 
\sin\theta$. Note, for comparison, that 
the one-loop-order chiral perturbation theory 
result is 1.022~\cite{gasser85b}. 
In regard to the $\sin 2\alpha$ results to follow, 
it is worth noting that the isospin-violating parameters we have
adopted appear conservative 
with respect to the existing experimental constraints.
Using $m_q(\mu=2.5\, {\rm GeV})$ of Ali {\it et al.}~\cite{ali98}, we find
$\varepsilon = 1.4 \cdot 10^{-2}$ and
$\varepsilon' = 7.7 \cdot 10^{-3}$; we use these values in
the subsequent calculations. 

\begin{figure}[ht]      
\centerline{\epsfxsize 7.0 truein \epsfbox{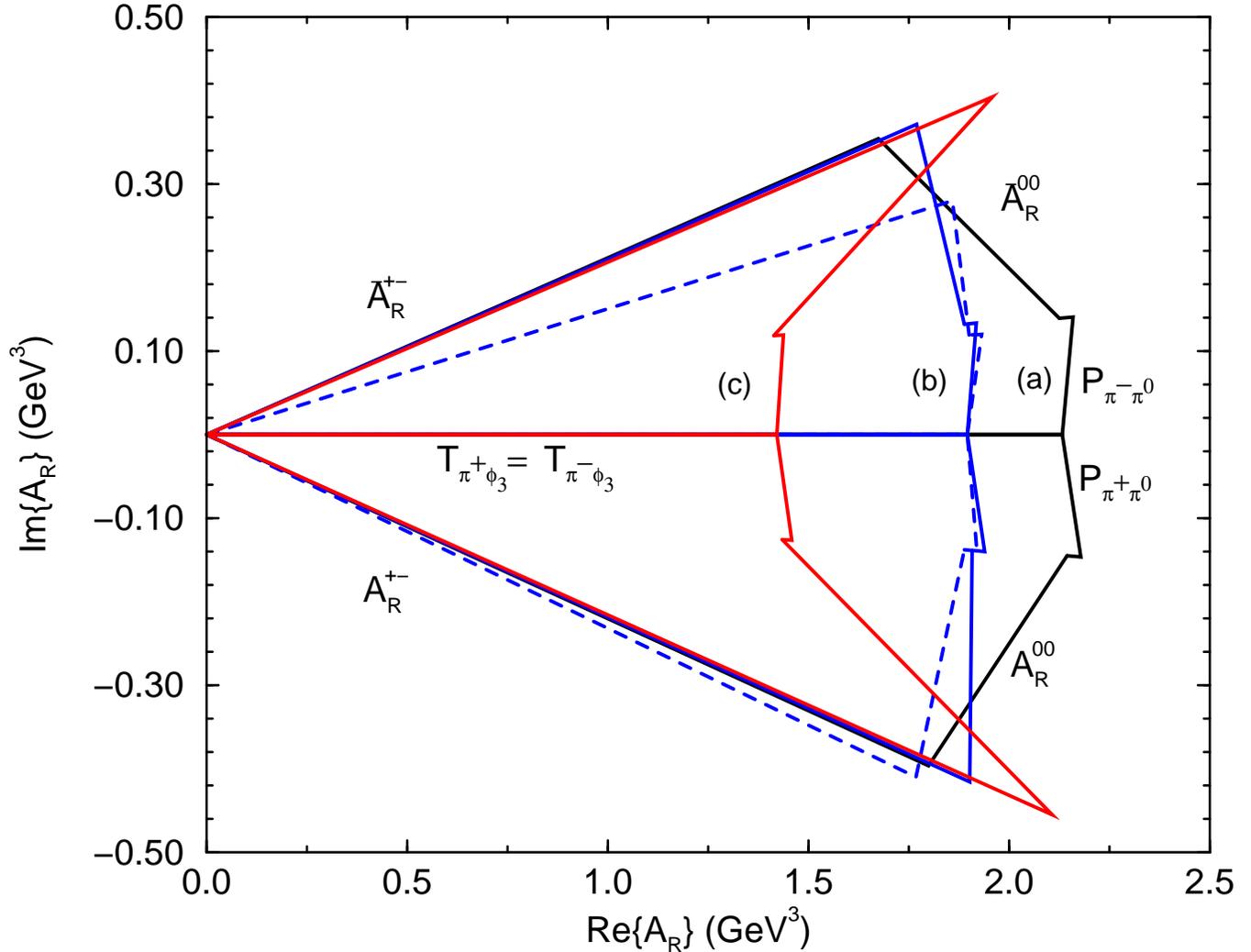}}
\vskip -.2 cm
\caption[]{
\label{figtri}
\small 
Reduced amplitudes 
in $B\ra\pi\pi$ 
in the factorization 
approximation with 
[$N_c$, $k^2/m_b^2$] for  
a) [2,0.5], b) [3, 0.5] (solid line) and [3, 0.3] (dashed line), and
c) [$\infty$, 0.5]. Note that 
${\overline A}_R^{\,00} \equiv 
2{\overline A}^{\,00} /((G_F/\sqrt{2}) i V_{ub} V_{ud}^*)$, 
${\overline A}_R^{\,+-} \equiv 
{\overline A}^{\,+-} /((G_F/\sqrt{2}) i V_{ub} V_{ud}^\ast)$, 
and 
$A_R^{-0} \equiv \sqrt{2}A^{-0} /((G_F/\sqrt{2}) i V_{ub} V_{ud}^\ast)$. 
The charged modes are separated into tree and penguin contributions,
so that 
$A_R^{+0}\equiv T_{\pi^+\phi_3} + P_{\pi^+\pi^0}$ and 
$A_R^{-0}\equiv T_{\pi^-\phi_3} + P_{\pi^-\pi^0}$, where 
$P_{\pi^\pm\pi^0}$ includes the isospin-violating tree
contribution in $A_R^{\pm0}$ as well. The shortest side in each
polygon is the vector defined by the RHS of Eq.~(\ref{newrel}); it is
non-zero only in the presence of \pieta mixing. 
}
\end{figure}

In the presence of \pieta mixing, the \Bpipi amplitudes
become 
%
\bea
 A^{-0}&=& \langle \pi^- \phi_3 | {\cal H}^{\rm eff} | B^-\rangle 
+ \varepsilon_8\langle \pi^- \phi_{8} | 
{\cal H}^{\rm eff} | B^-\rangle 
+ \varepsilon_0\langle \pi^- \phi_{0} | 
{\cal H}^{\rm eff} | B^-\rangle \\
{\overline A}^{\,00}&=& \langle \phi_3 \phi_3 | {\cal H}^{\rm eff} | 
{\bar B^0} \rangle 
+ \varepsilon_8\langle \phi_3 \phi_{8} | 
{\cal H}^{\rm eff} | {\bar B^0} \rangle 
+ \varepsilon_0\langle \phi_3 \phi_{0} | 
{\cal H}^{\rm eff} | {\bar B^0} \rangle \;,
\eea
\label{amps}
%
where $\varepsilon_0$ is the $\phi_3-\phi_0$ mixing angle,
$\varepsilon_0\equiv \varepsilon' \cos\theta - \varepsilon 
\sin\theta$. 
Note that either of the $\pi^0$ mesons in the $B^0\ra \pi^0\pi^0$ amplitude
can suffer \pieta mixing; the factor of two associated with this 
appears as $2{\overline A}^{\,00}$ in Eq.~(\ref{newrel}).
The $B\ra \pi\pi$ amplitudes satisfy
\bea
{\overline A}^{\,+-} + 2{\overline A}^{\,00} 
&-& \sqrt{2}\,A^{\,-0} 
= 
2\varepsilon_8 \langle \phi_3 \phi_{8} | 
{\cal H}^{\rm eff} | {\bar B^0}\rangle 
+2\varepsilon_0 \langle \phi_3 \phi_{0} | 
{\cal H}^{\rm eff} | {\bar B^0}\rangle \nonumber \\
&-& \sqrt{2} \varepsilon_8
\langle \pi^- \phi_{8} | 
{\cal H}^{\rm eff} | B^-\rangle 
- \sqrt{2} \varepsilon_0 
\langle \pi^- \phi_{0} | 
{\cal H}^{\rm eff} | B^-\rangle \;,
\label{newrel}
\eea
and thus the triangle relation of 
Eq.(~\ref{triangle}) becomes a quadrilateral.
We ignore 
the relatively unimportant mass
differences $m_{\pi^{\pm}}-m_{\pi^0}$ and $m_{B^{\pm}}-m_{B^0}$. 

\section{Results}

  We proceed by computing the individual amplitudes using the 
$\Delta B=1$ effective Hamiltonian resulting from the operator
product expansion in 
QCD in next-to-leading logarithmic (NLL) order~\cite{ali98}, using 
the factorization approximation for the hadronic matrix elements.
In this context, 
we can then apply the isospin analysis delineated above to
infer $\sin 2\alpha$ and thus estimate its theoretical systematic error, 
incurred through the neglect of isospin violating
effects. The effective Hamiltonian ${\cal H}^{\rm eff}$ 
for $b\ra d q\overline q$ decay 
can be parametrized as~\cite{ali98}
\be
{\cal H}^{\rm eff}= 
{G_F \over \sqrt{2}}\left[ V_{ub}V^*_{ud} (C_1 O_1^u + C_2 O_2^u)
+ V_{cb}V^*_{cd} (C_1 O_1^c + C_2 O_2^c)
 - V_{tb}V^*_{td} \left(\sum_{i=3}^{10} C_i O_i + C_g O_g\right) 
\right] \;,
\ee
where $O_i$ and $O_g$ are as per Ref.~\cite{ali98}; 
we also adopt their Wilson
coefficients $C_i$ and $C_g$, computed in the naive
dimensional regularization scheme at a renormalization scale 
of $\mu=2.5$ GeV~\cite{ali98}. 
In NLL order, the Wilson coefficients are scheme-dependent;
yet, after computing the hadronic matrix elements to one-loop-order,
the matrix elements of the effective Hamiltonian are still
scheme-independent~\cite{fleis93}. This 
can be explicitly realized
through the replacement 
$\langle dq{\overline q}| {\cal H}^{\rm eff} | b\rangle 
= (G_F/\sqrt{2}) \langle dq{\overline q} | 
[ V_{ub}V^*_{ud} (C_1^{\rm eff} O_1^u + C_2^{\rm eff} O_2^u)
 - V_{tb}V^*_{td} \sum_{i=3}^{10} C_i^{\rm eff} O_i ] | b \rangle^{\rm tree}$,
where ``tree'' denotes a tree-level matrix element and the 
$C_i^{\rm eff}$ are from Ref.~\cite{ali98}. 
The $C_i^{\rm eff}$ are complex~\cite{bss79} and 
depend on both the CKM matrix 
parameters
and $k^2$, where $k$ is the momentum transferred to the
$q\overline q$ pair in $b\ra d q\overline q$ decay. Noting Ref.~\cite{wolf83}
we use $\rho=0.12$, $\eta=0.34$, and $\lambda=0.2205$~\cite{ali98,SMparam}
unless otherwise stated. 
One expects $m_b^2/4 \lapp k^2 \lapp m_b^2/2$~\cite{kinem}; we 
use $k^2/m_b^2 = 0.3, 0.5$ in  what follows. 
We define the decay constants
$\langle \pi^- (p)| {\overline d} \gamma_\mu \gamma_5 u | 0 \rangle 
\equiv -i f_{\pi} p_{\mu}$ and 
$\langle \phi_i (p)| {\overline u} \gamma_\mu \gamma_5 u | 0 \rangle 
\equiv -i f_{\phi_I}^u p_{\mu}$, 
and use the flavor content of the SU(3)$_f$-perfect states
to relate $f_{\phi_i}^q$ to $f_{\pi}$.
Finally, using 
the quark equations of motion with PCAC and introducing
$a_i\equiv C_i^{\rm eff} + C_{i+1}^{\rm eff}/N_c$ for $i$ odd
and $a_i\equiv C_i^{\rm eff} + C_{i-1}^{\rm eff}/N_c$ for $i$ even,
the $B^-\ra \pi^-\phi_3$ 
matrix element in the factorization approximation with use of
the Fierz relations is
%
\setcounter{footnote}{0}
\vspace{3mm}
\begin{table}[delcf]
\begin{center}
\caption{Strong phases 
and inferred values of $\sin2\alpha$~\protect\cite{GL90} 
from amplitudes in the factorization
approximation with $N_c$ and $k^2/m_b^2=0.5$. 
The strong phase $2\delta_{\rm true}$
is the opening
angle between the ${\overline A_R}^{+-}$ and ${A_R}^{+-}$ amplitudes
in Fig.~1, 
whereas 2$\delta_{\rm GL}$ is the strong phase associated
with the closest matching $\sin2\alpha$ values, denoted 
$(\sin 2\alpha)_{\rm GL}$, from 
Im $r_{\pi^+\pi^-}$/Im $r_{\pi^0\pi^0}$, respectively. 
The bounds $|2\delta_{\rm GQI}|$ and $|2\delta_{\rm GQII}|$  on 
$2\delta_{\rm true}$ 
from Eqs.~(2.12) and (2.15) of Ref.~\protect\cite{GQpi97} are also shown.
All angles are in degrees.
We input a) $\sin2\alpha=0.0432$~\protect\cite{ali98,SMparam},
b) $\sin2\alpha=-0.233$ ($\rho=0.2,\eta=0.35$)~\protect\cite{DLcomm}, and 
c) $\sin2\alpha=0.959$ ($\rho=-0.12,\eta=0.34$).}
\begin{tabular}{ccccccc}

\hhha 
case & $N_c$ & 2$\delta_{\rm true}$ &  $|2\delta_{\rm GQI}|$ & 
$|2\delta_{\rm GQII}|$ &  $|2\delta_{\rm GL}|$ &  $(\sin 2\alpha)_{\rm GL}$  \\
\hline
a & 2 &  24.4 & 24.5 & 12.5 & 13.4 & -0.145/0.153 \\
a & 3 &  24.2 & 16.1 & 15.3  & 15.4  & -0.107/0.133 \\
a & $\infty$ & 23.8  & 55.8  & 19.4 & 17.7 & -0.0595/0.101 \\
b & 2 &  19.6 & 22.0 & 8.3 & 9.4 & -0.399/-0.0213 \\
b & 3 &  19.4 & 12.9 & 12.3  & 12.4  & -0.349/-0.0599 \\
b & $\infty$ & 19.2  & 56.2  & 17.3  & 15.7 & -0.287/-0.104 \\
c & 2 &  28.3 & 34.4 & 14.8 & 4.9 & 0.769/0.701 ($\ast,\dagger$)\\
c & 3 &  28.0 & 22.8 & 17.4  & 4.0  & 0.662/0.692 ($\ast,\dagger$)\\
c & $\infty$ & 27.6  & 42.7  & 21.2  & 19.5 & 0.912/0.967  
\end{tabular}
\label{delcf}
\end{center}
\setcounter{footnote}{0}
\vspace{-3mm}
\baselineskip=8 pt
{\footnotesize
$^\ast$ The matching
procedure fails to choose a $\sin2\alpha$ which is as 
close to the input value as possible. \\
$^\dagger$ The discrete ambiguity in the strong phase is resolved wrongly.}
\end{table}
\bea
&\langle& \pi^- \phi_3 | 
{\cal H}^{\rm eff} | B^-
\rangle 
= {G_F\over \sqrt{2}}
[ V_{ub}V^*_{ud} (i f_{\pi} F_{B^-\phi_3}(m_{\pi^-}^2)a_1
+ i f_{\phi_3}^u F_{B^-\pi}(m_{\pi^0}^2)a_2) 
- V_{tb}V^*_{td} \nonumber \\
&\times& 
(i f_{\pi} F_{B^-\phi_3}(m_{\pi^-}^2) 
(a_4 + a_{10} + {2m_{\pi^-}^2(a_6 + a_8)\over (m_u+m_d)(m_b - m_u)})
-
i f_{\phi_3}^u F_{B^-\pi}(m_{\pi^0}^2) 
\label{induced}
\\
&\times& 
(a_4 + {3\over 2}(a_7 - a_9) - {1\over 2}a_{10} 
+ {m_{\pi^0}^2(a_6 - {1\over 2}a_8)\over m_d(m_b - m_d)})
)]\;. \nonumber 
\eea
The transition form factors are given by
$F_{B^-\pi}(q^2)= 
(m_{B^-}^2 - m_{\pi^-}^2) F_0^{B\ra \pi}(0)/(1 - q^2/M_{0^+}^2)$,
where we use $F_0^{B\ra \pi}(0)=0.33$ and 
$M_{0^+}=5.73$ GeV as per Refs.~\cite{ali98,bsw87}. Also
$F_{B\phi_3}= F_{B\pi}/\sqrt{2}$,
$F_{B\phi_8}= F_{B\pi}/\sqrt{6}$, and
$F_{B\phi_0}= F_{B\pi}/\sqrt{3}$.
Note that the $a_4$ and $a_6$ terms, which are associated 
with the strong penguin operators, only contribute to 
the $\langle \pi^- \phi_3 | {\cal H}^{\rm eff} | B^-\rangle$ matrix
element if $m_u\ne m_d$ or if 
$f_{\phi_3}^u F_{B^-\pi}(m_{\pi^0}^2)\ne f_{\pi} F_{B^-\phi_3}(m_{\pi^-}^2)$ 
--- we neglect this latter contribution as we set
$m_{\pi^\pm}=m_{\pi^0}$. 
The strong-penguin contributions, which are isospin-violating, 
explicitly realize the induced $\Delta I=3/2$ effect
discussed previously, for the amplitude 
$\langle \pi^- \phi_3 | {\cal H}^{\rm eff} | B^-\rangle$, in concert
with the amplitudes
$\langle \pi^- \pi^+ | {\cal H}^{\rm eff} | \bar{B^0}\rangle$ 
and $\langle \phi_3 \phi_3 | {\cal H}^{\rm eff} | \bar{B^0}\rangle$,
satisfy the triangle relation of Eq.~(\ref{triangle}). 
Thus the $\langle \pi^- \phi_3 | {\cal H}^{\rm eff} | B^-\rangle$ 
amplitude
can still be deemed purely $I=2$ even if $m_u\ne m_d$, as
in Eq.~(\ref{induced}). 
In the presence of \pieta mixing, however, the 
$B^-\ra \pi^- \pi^0$ amplitude can no longer be purely 
$I=2$, as the RHS of Eq.~(\ref{newrel}) is non-zero and 
Eq.~(\ref{triangle}) is no longer satisfied. 

Numerical results 
for the reduced amplitudes $A_R$ and ${\overline A}_R$, where
${\overline A}_R^{\,00} \equiv 
2{\overline A}^{\,00} /((G_F/\sqrt{2}) i V_{ub} V_{ud}^*)$,
${\overline A}_R^{\,+-} \equiv 
{\overline A}^{\,+-} /((G_F/\sqrt{2}) i V_{ub} V_{ud}^\ast)$, and
$A_R^{-0} \equiv \sqrt{2}A^{-0} /((G_F/\sqrt{2}) i V_{ub} V_{ud}^\ast)$,
with $N_c=2,3,\infty$ and $k^2/m_b^2=0.3,0.5$ 
are shown in Fig.\ref{figtri}. 
$A_R^{+0}$ and $A_R^{-0}$ are
broken into tree and penguin contributions, so that 
$A_R^{+0}\equiv T_{\pi^+\phi_3} + P_{\pi^+\pi^0}$ and 
$A_R^{-0}\equiv T_{\pi^-\phi_3} + P_{\pi^-\pi^0}$. Note that
``$P_{\pi^\pm\pi^0}$'' is defined to include the isospin-violating tree
contribution in $A_R^{\pm0}$ as well. 
The shortest side in each
polygon is the vector defined by the RHS of Eq.~(\ref{newrel}).
The values of $\sin2\alpha$ extracted from the computed amplitudes 
with $N_c$ --- note that $N_c$ is 
regarded as an effective parameter in this context --- 
and $k^2/m_b^2=0.5$ are shown
in Table \ref{delcf}; the results for $k^2/m_b^2=0.3$ are similar
and have been omitted.
For reference, the
ratio of penguin to tree amplitudes in $B^-\ra \pi^-\pi^0$ is
$|P|/|T| \sim (2.2 - 2.7)\% | V_{tb}V^*_{td}| / |V_{ub}V^*_{ud}|$ for 
$N_c=2,3$ and $k^2$ as above. 
Were electroweak penguins the only source of isospin violation, then
$|P|/|T| \sim (1.4-1.5)\% | V_{tb}V^*_{td}| / |V_{ub}V^*_{ud}|$,
commensurate with the estimate of 1.6\% in Ref.~\cite{deshe95}. 

In the presence of $\pi^0$-$\eta,\eta'$ mixing,
the ${\overline A}_R^{\,+-}$, ${\overline A}_R^{\,-0}$, and 
${\overline A}_R^{\,00}$ amplitudes obey a quadrilateral relation 
as per Eq.~(\ref{newrel}).
Consequently, 
the values of $\sin2\alpha$ 
extracted from Im $r_{\pi^+\pi^-}$ and Im $r_{\pi^0\pi^0}$ 
measurements can not only differ
markedly from the value of $\sin 2\alpha$ input but also need not match.
The incurred error in $\sin 2\alpha$ increases as the value 
to be extracted decreases; the structure of Eq.~(\ref{rdef}) suggests this,
for as $\sin 2\alpha$ decreases, the quantity
Im $((1 - {\overline z})/(1 -z))$ becomes more important to 
determining the extracted value. 
It is useful to constrast the impact of 
the various isospin-violating effects.
The presence of $\Delta I=3/2$ penguin
contributions, be they from $m_u\ne m_d$ or electroweak effects, shift
the extracted value of $\sin 2\alpha$ from its input value, yet the 
``matching'' of the $\sin 2\alpha$ values from the
Im $r_{\pi^+\pi^-}$ 
and Im $r_{\pi^0\pi^0}$ determinations is
unaffected. This arises as the amplitudes in question still satisfy 
the triangle relations implied by Eq.~(\ref{triangle}). 
The mismatch troubles seen in Table \ref{delcf} are driven by
$\pi^0$-$\eta,\eta'$ mixing, though the latter shifts the values of
$\sin 2\alpha$ extracted from Im $r_{\pi^+\pi^-}$ as well. 
Picking the closest
matching values of $\sin 2\alpha$ in the two final states also 
picks the solutions closest to the input value; the exceptions are noted
in Table \ref{delcf}. The matching 
\setcounter{footnote}{0}
\vspace{3mm}
\begin{table}[alpherr]
\begin{center}
\caption{Errors in $\alpha$ were 
Im $r_{\pi^+\pi^-}$ taken to be $\sin 2\alpha$ 
($|\delta\alpha|_{\rm raw}$) and from applying the 
Gronau-London analysis~\protect\cite{GL90}
in the presence of isospin-violating corrections 
($|\delta\alpha|_{\rm GL}$) 
for amplitudes computed in the factorization 
approximation with $N_c=2$ and $k^2/m_b^2=0.5$. All angles
are in degrees. Cases a), b), 
and c) are defined as in Table I.}
\begin{tabular}{ccccccc}

\hhha 
case & $N_c$ & Im $r_{\pi^+\pi^-}$ & $|\delta\alpha|_{\rm raw}$ & $|\delta\alpha|_{\rm GL}$ \\
\hline
a & 2 &  -0.346 & 11.3 & 5.4 \\ 
b & 2 &  -0.514 & 8.8 & 5.0 \\ 
c & 2 &  0.642 & 16.8 & 11.6 ($\ast$) \\
\end{tabular}
\label{alpherr}
\end{center}
\setcounter{footnote}{0}
\vspace{-3mm}
\baselineskip=8 pt
{\footnotesize
$^\ast$ The discrete ambiguity in the strong phase is resolved wrongly
in this case --- see Table I.}
\end{table}
{\noindent 
procedure can also 
yield the wrong strong phase; in case c) of Table \ref{delcf} with
$N_c=2,3$, the triangles of the chosen solutions}
``point'' in the same direction, 
whereas they actually point oppositely. 
If $|A^{00}|$ and $|{\overline A}^{00}|$ 
are small~\cite{GL90} the complete isospin analysis may not be
possible, so that 
we also examine the utility of the bounds recently proposed by Grossman and
Quinn~\cite{GQpi97}
on the strong phase 
$2\delta_{\rm true}\equiv {\rm arg}((1 - {\overline z})/(1 - z))$
of Eq.~(\ref{rdef}). The bounds 
$2\delta_{\rm GQI}$ and $2\delta_{\rm GQII}$ given by their
Eqs.~(2.12) and (2.15)~{\cite{GQpi97}}, respectively, 
follow from Eq.~(\ref{triangle}), and thus can be broken
by isospin-violating effects. 
As shown in 
Table \ref{delcf}, the bounds typically are broken, and their 
efficacy does not improve as the value of $\sin 2\alpha$ 
to be reconstructed grows large.

  To conclude, we have considered the role of isospin violation
in $B\ra\pi\pi$ decays and have found the effects to be significant. 
Most particularly, the utility of the
isospin analysis in determining $\sin 2\alpha$ strongly depends 
on the value to be reconstructed. The error in $\sin 2\alpha$
from a Im $r_{\pi^+\pi^-}$ measurement grows markedly larger as
$\sin 2\alpha$ grows small --- this is the region of 
$\sin 2 \alpha$ currently favored, albeit weakly, by 
phenomenology~\cite{ali98,SMparam,DLcomm,mele98}. 
The effects found arise in part because the penguin contribution in 
$B \rightarrow \pi^+ \pi^-$, e.g., is itself small; we estimate
$|P|/|T| < 9\% | V_{tb}V^*_{td}| / |V_{ub}V^*_{ud}|$. Relative to this
scale, the impact of \pieta mixing is significant. This is displayed
in another way in Table~\ref{alpherr}. The ``penguin pollution'' 
in $B\ra \pi^+\pi^-$ is
such that were no isospin analysis applied, the error in $\alpha$
would be of the order of $10^\circ-20^\circ$. The isospin-violating effects
in $B\ra\pi^+\pi^-$ suggest that the error in $\alpha$ is still of the order
of $5^\circ$ after the 
Gronau-London~\cite{GL90} analysis is applied.
Yet, 
were the penguin
contributions in $B\rightarrow \pi\pi$ larger, pressing the need for
the corrections of the isospin analysis, 
the isospin-violating
effects considered would still be germane, for not only would 
the $\Delta I=3/2$ penguin contributions likely be larger, but 
the $B\rightarrow \pi\eta$ and
$B\rightarrow \pi\eta'$ contributions could also be 
larger as well~\cite{ciuchini97}. 
To conclude, we have shown that 
the presence of $\pi^0$-$\eta,\eta'$ mixing breaks the
triangle relationship, Eq.~(\ref{triangle}), usually assumed~\cite{GL90} 
and can mask the true value of $\sin2\alpha$.
%

%
%
\vspace{-3mm}
\section*{Acknowledgements}

I would like to thank 
Kam-Biu Luk and German Valencia 
for the opportunity to speak at this meeting. 
This work was supported by the U.S. Department of Energy 
under DE-FG02-96ER40989.

%
%

\end{document}